\documentclass[runningheads]{llncs}

\usepackage[english]{babel}
\usepackage[x11names]{xcolor} 
\usepackage{subcaption}
\usepackage{caption}[singlelinecheck=false]
\usepackage{url}
\usepackage{colortbl}
\usepackage{amssymb}
\usepackage{stmaryrd}
\usepackage{amsmath}
\usepackage{mathrsfs}
\usepackage{amssymb}
\usepackage[left]{lineno}
\usepackage{xfrac}
\usepackage{nicefrac}
\usepackage[textsize=scriptsize,backgroundcolor=yellow!40]{todonotes}
\usepackage[nonumberlist,acronym,sanitize=none]{glossaries}
\glsdisablehyper
\usepackage{comment}
\usepackage[pdftex, colorlinks=true, hyperfootnotes=true, hyperindex=true,
plainpages=false, pagebackref=false, pdfpagelabels=true, pdfstartview=FitH,
linkcolor=purple, citecolor=teal, urlcolor=blue,
bookmarks, bookmarksopen, bookmarksdepth=3]{hyperref}
\usepackage[capitalise,nameinlink]{cleveref}
\crefname{algocf}{Alg.}{Algs.}
\Crefname{algocf}{Algorithm}{Algorithms}
\crefname{aAlgocflLine}{line}{lines}
\Crefname{algocfline}{Line}{Lines}
\crefname{section}{Sect.}{Sects.} 
\Crefname{section}{Section}{Sections}
\crefname{table}{Tab.}{Tabs.} 
\Crefname{table}{Table}{Tables}
\usepackage{tkz-base}
\usetikzlibrary{decorations.pathmorphing,trees,snakes,arrows,shapes,automata,petri}
\usepackage{paralist}
\usepackage{multicol}
\usepackage{booktabs}
\usepackage[inline]{enumitem}
\usepackage{multirow}
\usepackage{rotating}
\usepackage{etaremune}
\usepackage[ruled,linesnumbered,noline,noend,algo2e]{algorithm2e}
\SetCommentSty{textrm} 
\SetArgSty{textrm} 
\SetFuncArgSty{textrm} 
\renewcommand{\SetProgSty}[1]{\renewcommand{\ProgSty}[1]{\textnormal{\csname#1\endcsname{##1}}\unskip}}
\SetProgSty{textrm} 
\usepackage{marginnote}
\usepackage{mathtools}
\usepackage{mathabx}
\usepackage{adjustbox}
\usepackage{ifthen}
\usepackage[normalem]{ulem}
\usepackage{lineno}
\usepackage{soul}
\usepackage{floatrow}
\usepackage{listings}
\crefname{lstlisting}{Listing}{Listings}
\crefname{lstfloat}{Listing}{Listings}
\usepackage{lipsum}
\usepackage{makecell}
\usepackage{diagbox}
\usepackage[scientific-notation=false,group-separator={,}]{siunitx}
\usepackage{microtype}
\usepackage{footmisc}
\usepackage{xspace}
\usepackage{ifdraft}
\usepackage{wrapfig}
\usepackage{scrextend}
\usepackage{orcidlink}
\usepackage{textcase}
\usepackage[ddmmyyyy]{datetime}

\newacronym{rdf}{RDF}{Resource Description Framework}
\newacronym{shacl}{SHACL}{Shapes Constraint Language}
\newacronym{sparql}{SPARQL}{SPARQL Protocol and RDF Query Language}
\newacronym{w3c}{W3C}{World Wide Web Consortium}
\newacronym{uri}{URI}{Uniform Resource Identifier}
\newacronym{owl}{OWL}{Web Ontology Language}
\newacronym{rml}{RML}{RDF Mapping language}
\newacronym{ttl}{Turtle}{Terse RDF Triple Language}
\newacronym{rdfs}{RDF/S}{RDF Schema}

\usepackage{newtxtext}

\newfloat{lstfloat}{htbp}{lop}
\floatname{lstfloat}{Listing}
\lstdefinelanguage{ocedttl}{%
	sensitive=false,%
	morekeywords={owl,rdf,xsd,rdfs,datetime,subClassOf,label,comment,domain,range,Class,DatatypeProperty,ObjectProperty,\@prefix},keywordstyle=\color{black!66},classoffset=9,%
	morekeywords={a},keywordstyle=\color{blue},classoffset=8,%
	morekeywords={res},keywordstyle=\setlength{\fboxsep}{1pt}\colorbox{gray!25},classoffset=7,%
	morekeywords={oced},keywordstyle=\setlength{\fboxsep}{1pt}\colorbox{blue!12},classoffset=6,%
	morekeywords={aux},keywordstyle=\setlength{\fboxsep}{1pt}\colorbox{yellow!25},classoffset=5,%
	morekeywords={ext},keywordstyle=\setlength{\fboxsep}{1pt}\colorbox{violet!25},classoffset=4,%
	morekeywords={event},keywordstyle=\setlength{\fboxsep}{1pt}\colorbox{pink!50},classoffset=3,%
	morekeywords={object},keywordstyle=\setlength{\fboxsep}{1pt}\colorbox{teal!24},classoffset=2,%
	morekeywords={ObjectRelation,ObjectAttribute},keywordstyle=\setlength{\fboxsep}{1pt}\colorbox{teal!6},classoffset=1,%
	morekeywords={res,observe,observe\_event,observed\_at,qualifier},keywordstyle=\setlength{\fboxsep}{1pt}\colorbox{gray!12},classoffset=0,%
	morecomment=[l]{\#},%
	morestring=[b]",%
	commentstyle=\color{gray},
	stringstyle=\color{brown},
}

\begin{document}
\title{\texorpdfstring{A Semantic Encoding of Object Centric Event Data}{A Semantic Encoding of OCED}}

\titlerunning{A Semantic Encoding of OCED}

\author{Saba~Latif\inst{1}\orcidlink{0000-0002-3989-8458}
        \and
        Fajar~J.~Ekaputra\inst{2}\orcidlink{0000-0003-4569-2496}
        \and
        Maxim~Vidgof\inst{2}\orcidlink{0000-0003-2394-2247}
        \and
        \\
        Sabrina~Kirrane\inst{2}\orcidlink{0000-0002-6955-7718}
        \and
		Claudio~Di~Ciccio\inst{3}\orcidlink{0000-0001-5570-0475}
}
\institute{
 	Sapienza University of Rome, Rome, Italy, \email{\href{mailto:saba.latif@uniroma1.it}{saba.latif@uniroma1.it}}
 	\and
	Wirtschaftsuniversität Wien, Vienna, Austria, \email{\href{mailto:fajar.ekaputra@wu.ac.at;maxim.vidgof@wu.ac.at;sabrina.kirrane@wu.ac.at}{firstname.lastname@wu.ac.at}}
    \and
	University of Utrecht, Utrecht, The Netherlands, \email{\href{mailto:c.diciccio@uu.nl}{c.diciccio@uu.nl}}
}

\authorrunning{S.\ Latif, F.J.\ Ekaputra, M.\ Vidgof, S.\ Kirrane, C.\ Di Ciccio}

\maketitle            

\begin{abstract}
The Object-Centric Event Data (OCED) is a novel meta-model aimed at providing a common ground for process data records centered around events and objects.
One of its objectives is to foster both interoperability and process information exchange.
In this context, the integration of data from different providers, the combination of multiple processes, and the enhancement of knowledge inference are novel challenges.
Semantic Web technologies can enable the creation of a machine-readable OCED description enriched through ontology-based relationships and entity categorization.
In this paper, we introduce an approach built upon Semantic Web technologies for the realization of semantic-enhanced OCED, with the aim to strengthen process data reasoning, interconnect information sources, and boost expressiveness. 
\end{abstract}

\section{Introduction}
\label{sec:intro}
Process mining is the discipline aimed at extracting, analyzing, and enhancing knowledge of business processes from event data stored by information systems in the form of event logs~\cite{vanderAalst2016}. 
Over the last few years, the focus of process mining has experienced a gradual drift from the historically established activity-centric view, interpreting process execution logs as sequences of actions~\cite{DBLP:conf/icpm/WynnLAACJV21}.
The spotlight is moving towards the information artifacts, namely \emph{objects}, that activity executions alter or read.
This viewpoint shift is testified by the surge of object-centric process mining~\cite{Aalst2019} and object-centric event log formats~\cite{Ghahfarokhi2021}. 

The IEEE Task Force on Process Mining leads the standardization process for a new event data paradigm, with the aim to overcome the previous established format of XES (eXtensible Event Stream,~\cite{DBLP:conf/icpm/WynnLAACJV21}): The Object Centric Event Data (OCED) meta-model~\cite{Fahland-etal/2024:OCED}.
An accurate semantic ontology of the novel meta-model is still under discussion. Ontologies function as explicit conceptual models representing domain knowledge, making it accessible to information systems. They are essential to the vision of the semantics, offering the semantic vocabulary needed to annotate websites so that machines can meaningfully interpret~\cite{GrimmAVS11}.

In this paper, we provide a framework that defines the semantics of OCED to endow the new standard with a consistent, extensible, and interoperable representation. We achieve this by leveraging the body of knowledge of the semantic web~\cite{BernersLee2001}. On top of that, we introduce an approach inspired by the literature in information integration~\cite{Calvanese.etal/CoopIS1998:InformationIntegrationConceptual} that resorts to semantic web description languages to establish links between the model at the meta-level (described by the OCED Ontology, OCEDO), customizable and domain-specific representations at an intentional level (via Domain-specific extensions of OCED, OCEDD), and the data at an extensional level (as knowledge graphs of OCED Resources, OCEDR).
We provide a demonstration of how raw process data can be automatically endowed with machine-readable semantics in an object-oriented fashion through a proof-of-concept prototype that automatically extracts OCEDR knowledge graphs from activity-centric XES logs. Our semantic framework encoding and tools are openly available at \href{https://github.com/wu-semsys/ocedo/}{\nolinkurl{github.com/wu-semsys/ocedo}}.

Our long-term goal is manifold. Among others, the foreseeable opportunities opened up by our approach range from an improvement of OCED data reasoning, to the knowledge inference from semantically enriched data, from an increase in expressiveness, to the interconnection of information sources. Regarding the last point, we remark that under our semantic umbrella, other (not necessarily process-specific) existing datasets can be linked with the resulting potential of unleashing unprecedented contextual enrichment for (semantic reasoning in) process mining.

The remainder of the paper is as follows. \Cref{sec:background} outlines the notions and technical bases of our investigation. \Cref{sec:ontology} proposes our three-layered conceptual framework for the semantic encoding of OCED. \Cref{sec:discussion} discusses the implications of our endeavor. \Cref{sec:relatedwork} overviews related work. \Cref{sec:conclusion} draws conclusions for this work.

\section{Background}
\label{sec:background}
%
%
Our investigation aims to forge semantic layers for object-centric event data using the OCED standard as the anvil, and the conceptual tools and frameworks provided by semantic web  technologies as the hammer. In this section, we provide a description of our conceptual forgery's items.

\subsection{The OCED Meta Model}
\label{sec:oced}
\begin{figure}[t]
	\includegraphics[width=.75\textwidth]{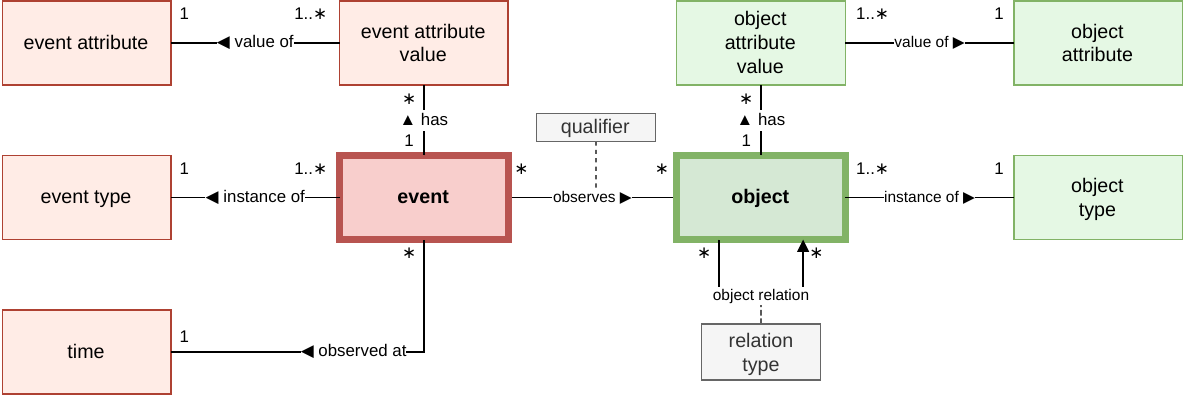}
	\caption[The OCED meta-model core]{The OCED meta-model core~\cite{Fahland-etal/2024:OCED}}
	\label{fig:metamodel}
\end{figure}

OCED aims to overcome the limitations of the well-established IEEE standard XES~\cite{IEEE2016}, which emerged in its ultra-decennial adoption at the time of writing. These limitations include lack of generalizability, complexity of the data structure, and memory expensiveness of the storage format~\cite{DBLP:conf/icpm/WynnLAACJV21}.
Unlike XES, OCED shifts the focus from sequences of activities recorded in information systems, trace after trace, to the lifecycle and conceptual interconnection of business objects that these systems handle.

A clear example of how semantically-aware object-centric formats can be useful to the representation of process execution records comes a public real-world event log tracing the process for incident and problem management at Volvo~\cite{Steeman/BPIChallenge2013}.
Although the event log is natively stored in XES, it comes endowed with several pieces of information pertaining to the treated business objects, and rich documentation for the data representation and contents.%
\footnote{\url{https://ais.win.tue.nl/bpi/2013/challenge.html}. Accessed: 26/09/2025.\label{ref:bpic2013}}
Furthermore, it demonstrates how raw, activity-centric event logs can be turned into semantically rich, object-centric information stores.
We shall use it as a running example in this paper, revisited under the lens of object-centric representations.

\Cref{fig:metamodel} graphically depicts the key elements of the OCED meta-model and their interrelation. 
Events and objects (marked with a thick bounding box in the figure) represent the first-class citizens of OCED.
An \textit{event} represents a point-in-time occurrence of an action. The \textit{objects} are entities of which the event may report the creation, change, deletion, or mere reading (i.e., that the object \textit{observes} in a way that the \textit{qualifier} clarifies).
For example, every status change of an incident (e.g., the one identified by the 
ticket number \texttt{1-364285768}) 
is an event (e.g., signaling that it closes the incident by acquiring a 
\texttt{Completed} status and \texttt{Closed} sub-status%
~\cite{Spiegel.etal/BPI2013:BPIC}). The reported incident, and hence the event, pertains to a product 
(e.g., \texttt{PROD582}).
The status change is operated by a responsible
(e.g., \texttt{Siebel})
of a support team
(e.g., \texttt{V5 3rd})
within an IT function division
(e.g., the one of \texttt{A2\_5}).
of a service center
(e.g., the one of \texttt{Org line A2}).
The incident, the product, the responsible, and the team are all objects in OCED.
Notice that there are relations between two objects in this example
(\texttt{Siebel} works in a team, which is part of an IT function division, which is in turn within a service center).
In OCED, the concept of \emph{object relation}, labelled by a \emph{relation type} (\textit{works in}, \textit{is part of} and \textit{is within}, in our example) bears this notion. 

Both events and objects can come endowed with attributes. In OCED, attributes are akin to name-value pairs; therefore, an object (like the incident) is associated to an \textit{object attribute value} for an \textit{object attribute} (e.g., \texttt{High} for the \texttt{Impact} of the problem), and an event (like the status change) is associated to an \textit{event attribute value} for an \textit{event attribute} (e.g., \textit{Completed} for the new incident's  \textit{status} and \textit{Closed} for its \textit{sub\_status}).
A \textit{time}-stamp is a special attribute since, unlike the other attributes, it has fixed semantics: In particular, it 
indicates when the event was registered (e.g., \texttt{2012-05-11T01:26:15+02:00}).

\subsection{Semantic Web Technologies}
\label{sec:semantic}
%
%
Semantic web technologies provide a comprehensive set of standards for the representation, linking, and processing of semantically explicit information.
For our intents and purposes, of particular relevance in the semantic web technology stack is the \gls{rdf}, which we use as a uniform model to represent object-centric event logs and the information they bear.
\Gls{rdf} is a data representation model published by the \gls{w3c}
as a set of recommendations and working group notes.%
\footnote{\url{http://www.w3.org/TR/rdf11-primer/}. Accessed: 30/09/2025.}
It provides a standard model for expressing information about \emph{resources}, which in our paper represent events, objects, attributes, relations, etc.
An \gls{rdf} dataset consists of a set of statements about these resources, expressed in the form of triples $(s,p,o)$ where $s$ is an RDF-\emph{subject}, $p$ is a \emph{predicate}, and $o$ is an RDF-\emph{object};
$s$ and $o$ represent the two resources being related whereas $p$ represents the nature of their relationship.
For instance, with \gls{rdf} we can declare that $\mathit{ev}_1$ (RDF-subject) is \textit{observed at} (predicate) \texttt{2012-05-11T01:26:15+02:00} (RDF-object).
A key advantage of \gls{rdf} as a data model is its extensible nature, i.e., additional statements about $s$, $p$, and $o$ can be added to link concepts and predicates from various additional, potentially domain-specific vocabularies.
To this end, the encoding of a resource is associated to a \gls{uri}, and grouped into namespaces that are used as prefixes to clarify the vocabulary they belong to.
The notion of RDF-object itself, for instance, is encoded as \texttt{rdf:object}, whereby the namespace \texttt{rdf} is associated to the \gls{uri} \texttt{http://www.w3.org/1999/02/22-rdf-syntax-ns\#}.
New vocabularies can be defined and connected. 

Notice that \gls{rdf} can cross the boundaries of information integration levels. The triple $(\mathit{ev}_1,\textit{observed at},\texttt{2011-02-03T08:28:58+01:00})$ is exerted on RDF-subject and RDF-object at the extensional level, whilst \textit{observed at} is a meta-model element.
We can also claim that $\mathit{ev}_1$ \textit{is a}n event, having the RDF-object at the meta-level.
The \gls{rdfs}%
\footnote{\Gls{rdfs}: \href{https://www.w3.org/TR/rdf-schema/}{\nolinkurl{www.w3.org/TR/rdf-schema}}. Accessed: 30/09/2025.}
is built upon the \gls{rdf} vocabulary and predicates over resources, classes categorizing resources, and the relations among classes.
For instance, with \gls{rdfs} we can express that \textit{observes} relates events (\emph{domain}) to objects (\emph{range}).
It is thus possible to automatically deduce that if $(\mathit{ev}_1, \textit{observes}, \mathit{obj}_1)$ is a declared triple, then $\mathit{ev}_1$ is an event and $\mathit{obj}_1$ an object.
Ontology specification languages such as the \gls{owl}%
\footnote{\Gls{owl}: \href{http://www.w3.org/TR/owl2-primer/}{\nolinkurl{www.w3.org/TR/owl2-primer}}. Accessed: 30/09/2025.}
can be used to more closely describe the semantic characteristics of terms in use. 
For example, with \gls{rdf} and \gls{owl} we can clarify that \textit{event} is a \texttt{owl:Class}, and that \textit{observed at} is an \texttt{owl:DataTypeProperty}.
\Gls{rdf} can be serialized via different formats, including RDF/XML, JSON-LD, and the text-based \gls{ttl}.%
\footnote{RDF/XML: \href{http://www.w3.org/TR/rdf-syntax-grammar/}{\nolinkurl{w3.org/TR/rdf-syntax-grammar}}; JSON-LD: \href{http://www.w3.org/TR/json-ld/}{\nolinkurl{w3.org/TR/json-ld}}; \gls{ttl}: \href{http://www.w3.org/TR/turtle/}{\nolinkurl{w3.org/TR/turtle}}. Accessed: 30/09/2025.}
Without loss of generality, we adopt \gls{ttl} in the remainder of the paper due to its compactness. 

%
A clear benefit of the aforementioned technologies is interoperability: information expressed in \gls{rdf} using shared vocabularies and ontologies can be exchanged between applications without loss of meaning. Furthermore, it makes it possible to apply a wide range of general purpose \gls{rdf} parsing, mapping, transformation, and query processing tools.
%
Once transformed into \gls{rdf}, information assumes the form of \emph{Knowledge Graphs} that can be published online, interlinked, and shared between applications and organizations, which is particularly interesting in the context of collaborative processes. 
Next, we will see how we use semantic web technologies as a building block for a semantic encoding of OCED that allows for an explicit stratification of domain-specific intensional layers over the meta-model.

\section{A Semantic Stratification of OCED}
\label{sec:ontology}
%
%
The OCED meta-model allows for flexibility in data representation, as it is meant to be \emph{domain-agnostic}. Notice, e.g., that no restrictions are exerted on
the possible relations that an object of a given type may or may not have with other objects,
the values that attributes can take,
or what (types of) objects are observed by which events.
For example, OCED does not natively provide users with means to enforce that each incident pertains to a product,
enumerate what the possible status changes can be,
or indicate that every status change event affects an incident.
It is also worth noticing that this meta-model removes a constraint that activity-centric standards like XES exerted, i.e., that every event be associated to a case, namely a process instance~\cite{DBLP:journals/ijcis/PourmirzaDG17,Bayomie.etal/IS2023:EventCaseCorrelation}. In the OCED meta-model, the case is not considered as a relevant, let alone mandatory, concept --however, it can be represented as an object.

While this structural looseness is desirable to encompass the amplest plethora of process data in a meta-model, domain experts should be allowed to inject knowledge on top of the meta-model, to provide additional details on the structure of the recorded data, promote traceability of process information provenance, and help the linkage of different data sources~\cite{Calvanese.etal/BIS2017:OntologyBasedData,Calegari.Delgado/BPMw2023:ModelDrivenEngineering,Piccirilli.etal/KI2025:PINPOINT}.
With this paper, we thus advocate the adoption of an overarching ontological approach to cover three conceptual layers, following a well-known approach in the field of information integration~\cite{Calvanese.etal/CoopIS1998:InformationIntegrationConceptual}: From the \emph{meta-level} (represented by the \emph{OCED O}ntology, OCEDO) to the \emph{intensional level} (the \emph{OCED D}omain-specific extension, OCEDD) to the \emph{extensional level} (the \emph{OCED R}esources' knowledge graph, OCEDR).
To define and interconnect the three levels, we resort to semantic web technologies, as we describe in the remainder of this section.

\subsection{The Meta-Level: OCED Ontology (OCEDO)}
\label{sec:ocedo}
\begin{figure}[tb]
	\centering
	\includegraphics[width=1\textwidth]{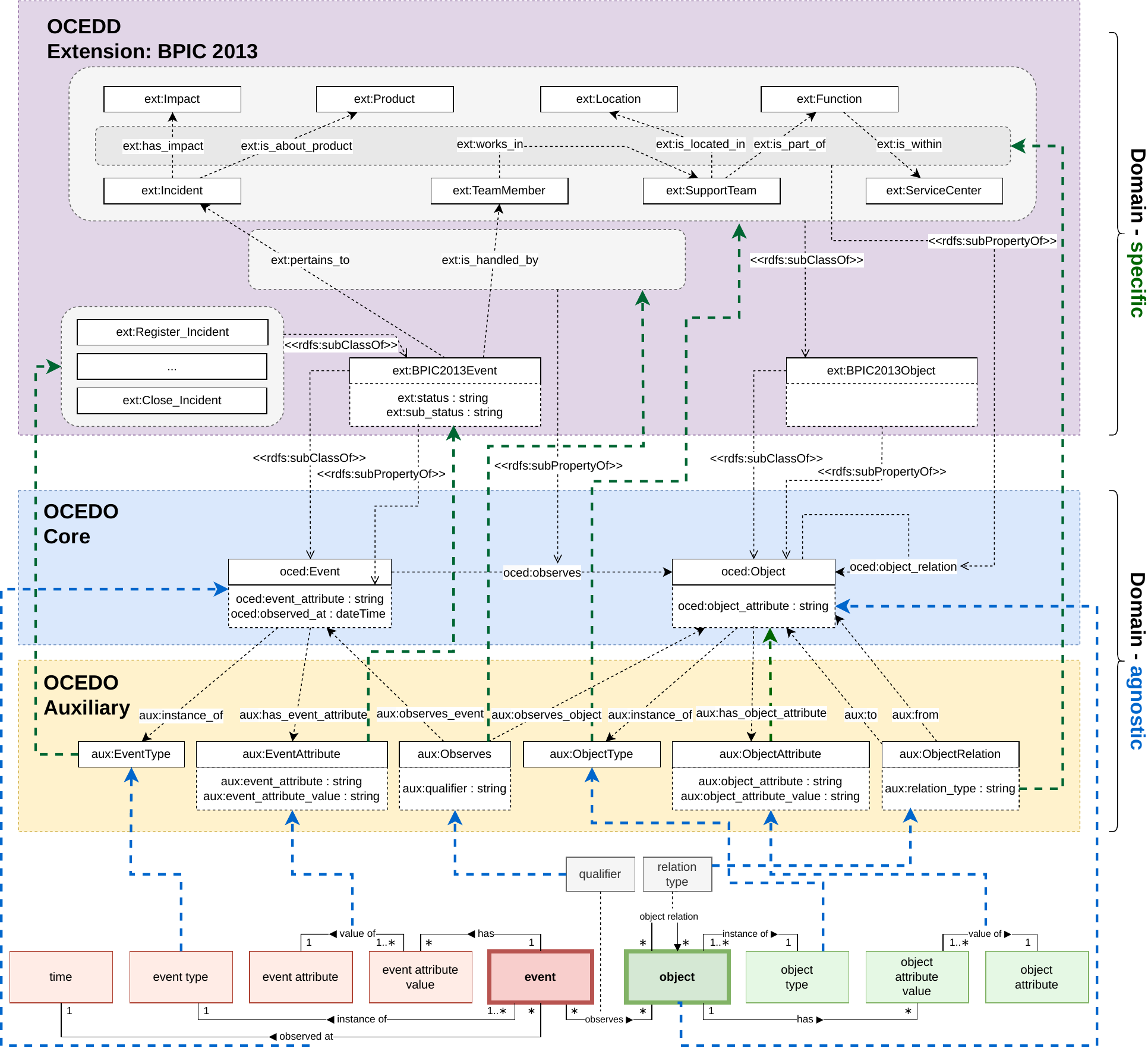}
	\caption{OCED meta-model with RDF representation and additional domain-specific layer}
	\label{fig:mm2ocedo2ocedd}
\end{figure}
\Cref{fig:mm2ocedo2ocedd} depicts our reference implementation of OCED's meta-model and intensional level in the form of an \gls{rdf} semantic network.
%
The OCED original meta-model entities and relations from \cref{fig:metamodel} occur below in the figure, as a reference. With dashed blue lines, we link those to the corresponding semantic encodings in our new OCED Ontology (OCEDO, in the middle). OCEDO is split in two parts. \emph{OCEDO Core} (prefixed with \texttt{oced}, namespace \url{https://w3id.org/ocedo/core\#}) encompasses the first-class citizens of OCED, namely \texttt{oced:Event} and \texttt{oced:Object}.%
\footnote{Notice that the prefix linked to \glspl{uri} allows for disambiguation: The notion of the \texttt{rdf:object}, defined by \gls{w3c}, does not match that of an \texttt{oced:Object}.} 
\emph{OCEDO Auxiliary} (\texttt{aux} prefix, namespace \href{https://github.com/wu-semsys/ocedo/blob/main/files/oced_ontology.ttl}{\nolinkurl{https://w3id.org/ocedo/aux\#}}) contains the representations of the other entities and relations in the original meta-model, namely the event type, event attribute, object type, object attribute, and the reified relations between objects, and from events to objects (\textit{observe}).

\begin{lstfloat}[tb]
\begin{lstlisting}[multicols=2,language=ocedttl]
@prefix rdf: <http://www.w3.org/1999/02/22-rdf-syntax-ns#> .
@prefix rdfs: <http://www.w3.org/2000/01/rdf-schema#> .
@prefix owl: <http://www.w3.org/2002/07/owl#> .
@prefix xsd: <http://www.w3.org/2001/XMLSchema#> .
@prefix oced: <https://w3id.org/ocedo/core#> .
@prefix aux: <https://w3id.org/ocedo/auxiliary#> .
# [...] OCED core
oced:Event a owl:Class ; § \label{line:ocedo:event} §
	rdfs:label "Event" ;
	rdfs:comment "Representation of the concept of Event" .
oced:Object a owl:Class ;  § \label{line:ocedo:object} §
	rdfs:label "Object" ;
	rdfs:comment "Representation of the concept of Object" .
# [...] OCED core - data properties
oced:observed_at a owl:DatatypeProperty ; § \label{line:ocedo:observed-at:s} §
	rdfs:label "observed_at" ;
	rdfs:comment "timestamp of an event" ;
	rdfs:domain oced:Event ;
	rdfs:range xsd:dateTime . § \label{line:ocedo:observed-at:e} §
# [...] OCED auxilary - classes
aux:Observe a owl:Class ; § \label{line:ocedo:observe} §
	rdfs:label "Observe" .
aux:ObjectAttribute a owl:Class ; § \label{line:ocedo:objectattribute} §
	rdfs:label "Object Attribute" .
aux:ObjectRelation a owl:Class ; § \label{line:ocedo:objectrelation} §
	rdfs:label "Object Relation" .
# [...] OCED auxilary - object properties
aux:has_object_attribute a owl:ObjectProperty ;   § \label{line:ocedo:observe:has-object-attribute:s} §
	rdfs:label "has_object_attribute" ;
	rdfs:domain oced:Object ;
	rdfs:range aux:ObjectAttribute . § \label{line:ocedo:observe:has-object-attribute:e} §
aux:from a owl:ObjectProperty ; § \label{line:ocedo:observe:from:s} §
	rdfs:label "from" ;
	rdfs:domain aux:ObjectRelation ;
	rdfs:range oced:Object ;
	rdfs:comment "source Object" . § \label{line:ocedo:observe:from:e} §
aux:to a owl:ObjectProperty ; § \label{line:ocedo:observe:to:s} §
	rdfs:label "to" ;
	rdfs:domain aux:ObjectRelation ;
	rdfs:range oced:Object ;
	rdfs:comment "target Object" . § \label{line:ocedo:observe:to:e} §
aux:observe_object a owl:ObjectProperty ;  § \label{line:ocedo:observe:object:s} §
	rdfs:label "observe_object" ;
	rdfs:domain aux:Observe ;
	rdfs:range oced:Object ;
	rdfs:comment "Relation between Observe and related Object" . § \label{line:ocedo:observe:object:e} §
aux:observe_event a owl:ObjectProperty ;  § \label{line:ocedo:observe:event:s} §
	rdfs:label "observe_event" ;
	rdfs:domain aux:Observe ;
	rdfs:range oced:Event ;
	rdfs:comment "Relation between Observe and related Event" . § \label{line:ocedo:observe:event:e} § # [...]
\end{lstlisting}
	\caption{A \acrshort{ttl} representation of the OCEDO ontology (excerpt)}
	\label{lst:ocedo}
\end{lstfloat}
Here we provide a brief overview of the OCED meta-model's encoding with semantic web technologies. A full specification is available in our open code repository.%
\footnote{\href{https://semsys.ai.wu.ac.at/ocedo/}{\nolinkurl{https://semsys.ai.wu.ac.at/ocedo}}. Accessed: 30/09/2025.\label{ref:repo}}
\Cref{lst:ocedo} shows an excerpt of the OCEDO schema serialized in \gls{ttl}. In the listing, we colored the background of keywords compatibly with the color scheme of \cref{fig:mm2ocedo2ocedd} for understandability purposes.
After the import of existing ontologies via prefixes and namespaces (like the aforementioned \gls{rdf}, \gls{rdfs}, \gls{owl}, see \cref{sec:semantic}) we declare that what is preceded by \texttt{oced:} and \texttt{aux:} identifies concepts for our core and auxiliary ontology parts of the meta-model, respectively.
Thereafter, starting from line~\ref{line:ocedo:event}, we state that \texttt{oced:Event} is a class to be \texttt{label}ed ``Event'', to be regarded as a representation of the event class (see the \texttt{comment} directive; both \texttt{label} and \texttt{comment} are terms of the \texttt{rdfs} vocabulary). Similarly, we introduce the concept of \texttt{oced:Object} (line~\ref{line:ocedo:object}). 

Among the concepts of the auxiliary section of OCEDO, we report here that of \texttt{aux:ObjectAttribute}, and of the reified relation types  \texttt{aux:ObjectRelation} and \texttt{aux:Observe} (see lines~\ref{line:ocedo:objectattribute},~\ref{line:ocedo:objectrelation}~and~\ref{line:ocedo:observe}, respectively).
Notice that the mappings from objects to their attributes, from objects to the relation to other objects, and from events to the observation of objects are all defined in the form of an \texttt{owl:ObjectProperty}, specifying the \texttt{rdfs:domain} and \texttt{rdfs:range} of such mappings (see \cref{sec:semantic}), as can be noticed on lines~\ref{line:ocedo:observe:has-object-attribute:s}-\ref{line:ocedo:observe:has-object-attribute:e}, \ref{line:ocedo:observe:from:s}-\ref{line:ocedo:observe:to:e}, and \ref{line:ocedo:observe:object:s}-\ref{line:ocedo:observe:event:e}.
Since the timestamp of an event is a scalar value of a type provided by the well-established XML-schema (\texttt{xsd}) vocabulary, we indicate that \texttt{observed\_at} is an \texttt{owl:DataTypeProperty} and specify that the \gls{rdfs} domain and range of it are the \texttt{oced:Event} class and the \texttt{xsd:dateTime} datatype, respectively (see the block starting on lines~\ref{line:ocedo:observed-at:s}-\ref{line:ocedo:observed-at:e}).

\subsection{The Intensional Level: Domain-specific Extension (OCEDD)}
\label{sec:ocedd}
OCEDO is domain-agnostic. Regardless of the organization registering the process data, and the structure thereof, it does not vary.
The members of the OCED Working Group encourage work for enriching process data semantics with domain-specific knowledge in their white paper~\cite{Fahland-etal/2024:OCED}.
To cater for this, we leverage the extensibility of \gls{rdf} to postulate the addition of another semantic layer at the intensional level: The domain-specific extensions which we collectively name as OCEDD (henceforth prefixed with \texttt{ext}).
The abstraction step is akin to the passage from the notion of entity and relationship types as modeling concepts to the conceptual schema of a relational database, with actual entities and relationships representing the business domain of data. Notice, however, that our modeling objective is descriptive and not normative like an ER diagram.

While OCEDO is meant to encode the semantics of the OCED meta model based on the directives of the XES/OCED working group~\cite{Fahland-etal/2024:OCED}, OCEDD is intended as a customizable addendum to create dialects of OCED that are domain- or case-specific. OCEDD extensions allow domain experts, analysts, but also data mining tools, to enrich the information brought by recorded process data with additional knowledge tailored for the situation in use in a controlled fashion, thus propelling \emph{extensibility}. However, OCEDD classes are intended to extend the existing, domain-agnostic OCEDO counterparts, so as to foster retro-compatibility and \emph{interoperability}. 

\begin{lstfloat}[tb]
\begin{lstlisting}[multicols=2,language=ocedttl]
@prefix ext: <https://w3id.org/ocedo/domain#> .
# [...] Domain-specific event and object types
ext:BPIC2013Event a owl:Class ; § \label{line:ocedd:bpic2013event} §
	rdfs:subClassOf oced:Object .
ext:BPIC2013Object a owl:Class ; § \label{line:ocedd:bpic2013object} §
	rdfs:subClassOf oced:Object .
ext:Incident a owl:Class ; § \label{line:ocedd:incident} §
	rdfs:subClassOf ext:BPIC2013Object .
ext:TeamMember a owl:Class ; § \label{line:ocedd:teammember} §
	rdfs:subClassOf ext:BPIC2013Object .
ext:SupportTeam a owl:Class ; § \label{line:ocedd:supportteam} §
	rdfs:subClassOf ext:BPIC2013Object .
ext:Completed_Resolved a owl:Class ; § \label{line:ocedd:completed-resolved} §
	rdfs:subClassOf ext:BPIC2013Event .
# [...] Domain-specific event-object and object-object relations
ext:pertains_to a owl:ObjectProperty ; § \label{line:ocedd:pertains-to:s} §
	rdfs:label "ext:pertains_to" ;
	rdfs:domain oced:Event ;
	rdfs:range ext:Incident . § \label{line:ocedd:pertains-to:e} §
ext:is_handled_by a owl:ObjectProperty; § \label{line:ocedd:is-handled-by:s} §
	rdfs:label "ext:is_handled_by" ;
	rdfs:domain oced:Event ;
	rdfs:range ext:TeamMember . § \label{line:ocedd:is-handled-by:e} §
ext:works_in a owl:ObjectProperty ; § \label{line:ocedd:works-in:s} §
	rdfs:label "ext:works_in" ;
	rdfs:domain ext:TeamMember ;
	rdfs:range ext:SupportTeam . § \label{line:ocedd:works-in:e} §
# [...] Domain-specific event-attribute relations
ext:status a owl:DatatypeProperty ; § \label{line:ocedd:status:s} §
	rdfs:subPropertyOf oced:event_attribute . § \label{line:ocedd:status:e} §
ext:substatus a owl:DatatypeProperty ; § \label{line:ocedd:substatus:s} §
	rdfs:subPropertyOf oced:event_attribute . § \label{line:ocedd:substatus:e} §
\end{lstlisting}
	\caption{A \acrshort{ttl} representation of an OCEDD extension for BPIC 2013 (excerpt)}
	\label{lst:ocedd}
\end{lstfloat}
Let us consider again the example of BPIC 2013.
Some concepts therein solely pertain to the domain-specific nomenclature of the terms (like incidents, products, responsible person, support teams) and to the constraints binding the concepts the information system reports on (every incident pertains a product, and every status update thereof is handled by a responsible person who works in a support team). Therefore, they do not apply to OCED as a whole, but to the specific extension thereof to use in this case.
In what follows, we provide an example of a possible domain-specific encoding of the BPIC~2013 data linked with OCEDO. Our reconstruction is tentative and based on the dataset documentation%
\textsuperscript{\ref{ref:bpic2013}}
and the analysis of the challenge submissions like~\cite{Spiegel.etal/BPI2013:BPIC} for exemplification purposes. It is by no means intended to be comprehensive or devoid of interpretation mistakes, but serves the purpose of clarifying the rationale of OCEDD.

\Cref{lst:ocedd} shows an excerpt of an OCEDD extension for the BPIC2013 log. An overview of it is graphically depicted in the topmost layer of \cref{fig:mm2ocedo2ocedd}, with dashed green arcs highlighting some implicit connections between OCEDD elements and the related OCEDO concepts below whenever entailed, but not explicitly enforced, by \gls{rdf} primitives. The full encoding is publicly available in our codebase.%
\textsuperscript{\ref{ref:repo}}
Here we focus on a few salient characteristics.
We begin defining a sub-class of \texttt{oced:Event} and \texttt{oced:Object} to characterize the concepts of pertinence for the specific log, here \texttt{ext:BPIC2013Event} (line~\ref{line:ocedd:bpic2013event}) and \texttt{ext:BPIC2013Object} (line~\ref{line:ocedd:bpic2013object}), respectively.
Event types categorize the possible \texttt{ext:BPIC2013Event}s via \texttt{rdfs:subClassOf} (see line~\ref{line:ocedd:completed-resolved} for the status transition of the incident to being completed and resolved). 
Objects of interest in this domain include \texttt{ext:Incident}, \texttt{ext:TeamMember}, and \texttt{ext:SupportTeam} (the names recall the ones mentioned in \cref{sec:oced}; see lines~\ref{line:ocedd:incident}, \ref{line:ocedd:teammember} and \ref{line:ocedd:supportteam} in \cref{lst:ocedd}).

\begin{sloppypar}
In our example scenario, all events are bound to an incident. We express this concept by means of the \texttt{owl:ObjectProperty} we adopted in OCEDO (see \cref{sec:ocedo}): thereby, the \texttt{pertains\_to} relationship binds \texttt{oced:Event}s, the domain, to an \texttt{ext:Incident} (range; see lines~\ref{line:ocedd:pertains-to:s}-\ref{line:ocedd:pertains-to:e}). Similarly, we indicate that an \texttt{oced:Event} is handled by an \texttt{ext:TeamMember} (lines~\ref{line:ocedd:is-handled-by:s}-\ref{line:ocedd:is-handled-by:e}),and that an \texttt{ext:TeamMember} works in an \texttt{ext:SupportTeam} (lines~\ref{line:ocedd:works-in:s}-\ref{line:ocedd:works-in:e}).
Through the aforementioned \texttt{owl:DatatypeProperty}, we can associate events to given attributes like the \texttt{ext:status} and \texttt{ext:substatus} by declaring those as an \texttt{rdfs:subPropertyOf} of the \texttt{oced:event\_attribute} (see lines~\ref{line:ocedd:status:s}-\ref{line:ocedd:status:e} and \ref{line:ocedd:substatus:s}-\ref{line:ocedd:substatus:e}, respectively).
\end{sloppypar}

\subsection{The Extensional Level: Resources' (OCEDR) Knowledge Graphs}
\label{sec:ocedr}
Equipped with the description of the domain-specific information about events and objects in the process data, we proceed with the third and last, extensional level. To this end, we introduce the OCEDR layers.

\begin{figure}[tb]
	\centering	\includegraphics[width=.75\textwidth]{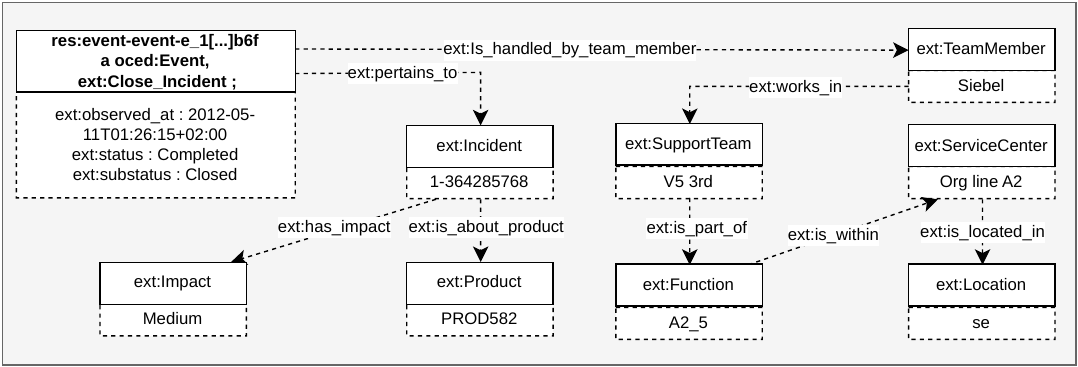}
	\caption{A section of the OCEDD knowledge graph from BPIC2013}
	\label{fig:BPIC2013:KGexample}
\end{figure}
\begin{lstfloat}[tb]
\begin{lstlisting}[multicols=2,language=ocedttl]
@prefix res: <https://w3id.org/ocedo/resource/> .
# [...] An event
res:event-e_1[...]b6f a oced:Event, ext:Close_Incident ; § \label{line:ocedr:event:s} §
		oced:observed_at "2012-05-11T01:26:15+02:00"^^xsd:dateTime ;
		ext:status "Completed" ;
		ext:pertains_to res:object-o_5[...]374 ;
		ext:is_handled_by res:object-o_e[...]011 ;
		ext:substatus "Closed" . § \label{line:ocedr:event:e} §
# [...] An object of type Incident
res:object-o_5[...]374 a oced:Object, ext:Incident ; § \label{line:ocedr:incident} §
	ext:ticket_number "1-364285768" .
	ext:is_about_product res:object-o_7[...]846 ; § \label{line:ocedr:is-about-product} §
# [...] An object of Type TeamMember
res:object-o_e[...]011 a oced:Object, ext:TeamMember ; § \label{line:ocedr:teammember} §
	ext:team_member_name "Siebel" ;
	ext:works_in res:object-o_8[...]430, § \label{line:ocedr:works-in} §
# [...] An object of Type SupportTeam
res:object-o_b[...]430 a oced:Object,
	ext:SupportTeam ; § \label{line:ocedr:supportteam} §
	ext:team "V5 3rd" ;
# [...] An object of type Product
res:object-o_7cb5c41b1041559e8f4b2d1286c58459145ac846 a oced:Object, ext:Product ; § \label{line:ocedr:product} §
	ext:product_number "PROD582" .
\end{lstlisting}
	\caption{A \acrshort{ttl} representation of an OCEDR knowledge graph for the BPIC 2013 datastore (excerpt)}
	\label{lst:ocedr:ttl}
\end{lstfloat}
\Cref{fig:BPIC2013:KGexample} shows an excerpt of the BPIC2013 event log turned into an OCEDR knowledge graph, revolving around a specific event from our running example in \cref{sec:oced}. In \cref{lst:ocedr:ttl}, we show a fragment of the corresponding \gls{ttl} encoding. In the latter, we refer to the entries instantiating the concepts in OCEDO and OCEDD generically as resources (prefixed with \texttt{res}). On line~\ref{line:ocedr:event:s}, e.g., we indicate that \texttt{event-e\_1[...]b6f} is an event of type \texttt{ext:Close\_Incident}. In the indented lines that follow (till \ref{line:ocedr:event:e}), we specify its status and substatus attributes, its timestamp, and the relations to two more resources instantiating objects: among others, the \texttt{ext:Incident} it pertains to (namely the one with ticket number \texttt{1-364285768}, see line~\ref{line:ocedr:incident}), and the \texttt{ext:TeamMember} (whose name attribute is \texttt{Siebel}, see line~\ref{line:ocedr:teammember}). Some relations among objects are shown on lines~\ref{line:ocedr:works-in} (to indicate that Siebel works in team \texttt{V5 3rd}, see line~\ref{line:ocedr:supportteam}) and~\ref{line:ocedr:is-about-product} (to indicate that the incident pertains to product number \texttt{PROD582}, see line~\ref{line:ocedr:product}).

Thus far, we have discussed with examples the core concepts underpinning our three-layered conceptual framework for the semantic description of OCED data. Next, we discuss ongoing work and future implications for its possible adoption and enhancement.

\section{Potential Impact and Future Work}
\label{sec:discussion}
Our proposed framework is a first building block, on which diverse research endeavors can be based.
The core contribution here is the systematization of semantic object-centric process data modeling with distinct strata: Meta-level, intensional level, and extensional level.
To generate the examples in this paper, and have a preliminary proof-of-concept testing our framework, we created a tool for the automated extraction of OCEDR knowledge graphs out of flattened event logs in XES format. It is openly available at \href{https://github.com/wu-semsys/ocedo/}{\nolinkurl{github.com/wu-semsys/ocedo}}. It takes as input an XES event log, an OCEDD document, and a tabular descriptor linking XES entry names to OCEDO and OCEDD concepts. The descriptor, e.g., dictates that the \texttt{org:resource} and \texttt{org:group} XES event attributes respectively contain the values of the \texttt{name} attribute of an \texttt{ext:TeamMember} in \texttt{ext:is\_handled\_by} relation with that \texttt{oced:Event}, and the \texttt{team} attribute of an \texttt{ext:SupportTeam} object in \texttt{ext:works\_in} relation with the latter. The tool leverages \gls{rml}%
\footnote{\href{https://rml.io/specs/rml/}{\nolinkurl{https://rml.io/specs/rml/}}. Accessed: 30/09/2025} for the mapping from input to output resources.
Further tests conducted with other real-world event logs in the financial and governmental domains can be found there.
A description of the tool goes beyond the scope of this paper. However, it serves as a clarion call for further implementations that link not only XES but other formats including raw CSVs, relational databases, OCEL files to OCEDR graphs. The more the connectors, the higher the impact that a semantic framework can yield. Furthermore, they can allow for the linkage of heterogeneous data repositories from multiple sources, enhancing knowledge inference capabilities, and enriching expressiveness. Information integration is indeed a driving factor that motivates our work.

The idea underneath our proof of concept takes inspiration from noticeably sophisticated approaches proposed for the automated extraction of activity-centric event logs like ONPROM~\cite{Calvanese.etal/EKAW2018:ONPROM}. Our vision is that the transfer of knowledge from those notable endeavors to the object-centric paradigm intertwined with our semantic encoding can unleash unprecedented results.
Among those, we foresee a novel holistic mining approach: While object-centric process mining stands out as the natural extension towards information extraction from OCEDR knowledge graphs, an intriguing direction points at the use of machine learning approaches to automatically infer event-to-object and object-to-object relations based on raw event data.
It is also worth mentioning that \gls{rdf} datastores cater for query engines that retrieve, aggregate and manipulate semantically rich information via \gls{sparql},%
\footnote{\url{https://www.w3.org/TR/sparql11-query/}. Accessed: 30/09/2025.}
opening up the opportunity for an advanced intermediate layer between data and process mining algorithms.

While the creation of OCEDD documents can be a way to encode the knowledge of knowledge experts, visual languages and model-driven approaches should be developed so as to allow their definition without necessarily being fluent in \gls{rdf} and derivatives.
Furthermore, OCEDD documents could be complemented with normative specifications dictating the structure of process data (e.g., dictating that every team member \emph{must be} associated to exactly one support team). To this end, existing standardized languages like \gls{shacl} could come handy (a first attempt in this direction was shown in~\cite{DiCiccio2019} for compliance checking). Reasoning on description and prescription documents looking for redundancies and contradictions, as well as to deduce connections among concepts, is a core task for semantic web and the foundational track of process management.

\section{Related Work}
\label{sec:relatedwork}
Our investigation relates with the semantic enrichment of process data, and to object-centric approaches to process mining. Next, we overview some notable pieces of work investigating those two areas.

The efforts to standardize an exchange format for event logs began in 2003%
\footnote{Quoting the words of the initiator, Wil van der Aalst, on the occasion of the 10th anniversary of XES: \url{https://www.tf-pm.org/newsletter/newsletter-stream-4-12-2020/10-years-of-xes}. Accessed: 28/09/2025.}
with the XML-based format MXML (Mining eXtensible Markup Language)~\cite{Dongen.Aalst/EMOIINTEROP2005:MXML}.
The well-known open-source process mining toolkit ProM~\cite{Aalst.etal/BPMdemos2009:ProMProcessMining} used MXML as the language of choice for input event logs.
In their paper, the authors postulated key concepts for process mining data treatment: The instantaneous nature of an event, without duration, the typing of events to classify the activity they report on, the association of data to a process, and the association of a process instance (case) to an event.
Soon after, Alves~de~Medeiros et al.~\cite{AlvesdeMedeiros.etal/ECIS2008:SAMXML} advocated the need for a semantic stratum to lay on data to allow mining techniques to consolidate, link and reason on concepts rather than strings. Hence their proposal: Semantically Annotated Mining eXtensible Markup Language (SA-MXML), annotating MXML entries with ontology references. We share their goal of semantically enriching the notions included in an event log, although we aim to operate at a deeper level by representing the whole standard as a multi-layered extensible knowledge graph.
Bertoli et al.~\cite{Bertoli2022} use an OWL knowledge base to represent business processes with semantic annotations. They advance prior research by encoding data artifacts, integrating execution traces, and enhancing semantic modeling. Furthermore, they improve on collaborative modeling and execution analysis using semantic reasoning techniques~\cite{Kampik2022}. The cross-fertilization of those disciplines has already brought about fruitful results (cf.~\cite{%
DiCiccio2019,Baumann2023,Bachhofner2022}), which we seek to further spur. 

Object-centric process mining has been gaining traction in recent years~\cite{Adams2022}.
Several approaches to object-centric process mining have been proposed, by enhancing techniques for imperative~\cite{Li2018,DBLP:journals/fuin/AalstB20} and declarative languages~\cite{christfort2024discovery}, or for completely new paradigms like the Object-Centric Behavioral Constraints (OCBC)~\cite{Artale2019,DBLP:conf/bis/LiCA17}. Interestingly, the notion of knowledge graph was presented in the process mining community for a more comprehensive view on multi-dimensional analyses, covering control flow, data flow, resources, and time, with the event knowledge graphs~\cite{DBLP:books/sp/22/Fahland22}.
By adding a semantic layer to the event data network introduced thereby, we approach can be integrated with that.

Before and concurrently to OCED, other object-centric event log languages have been introduced, such as eXtensible Object-Centric (XOC) logs~\cite{Li2018}, and Object-Centric Event Logs (OCEL)~\cite{Ghahfarokhi2021}.
Especially OCEL has observed a production of several utilities to handle and process it, including a visualization tool~\cite{Ghahfarokhi2022}, case and variant comparison~\cite{Adams2022}, filtering and sampling~\cite{Berti2022}. Our framework could amply leverage this body of knowledge to promote semantic web reasoning tasks on OCED datastores. 


\section{Conclusion}
\label{sec:conclusion}
Recent advances in object-centric process mining underscore the need for a formal semantic definition of Object-centric Event Data (OCED) to ensure consistency, extensibility, and interoperability. By leveraging semantic web technologies, we presented a machine-readable semantic framework that enhances ontology-based relationships and entity categorization for OCED. Our approach involves a three-level knowledge representation, having OCEDO at the meta-level, OCEDD domain-specific extensions at the intensional level, and OCEDR knowledge graphs at the extensional level. 
We demonstrated how our framework can be employed to automatically endow raw event logs with data semantics, with the help of a proof-of-concept tool we implemented to this end.
Through iterative refinements, we seek to build upon the presented results and deliver new techniques to improve OCED data reasoning, link heterogeneous data sources, and enhance knowledge inference and expressiveness.

\bibliographystyle{splncs04}
\bibliography{bib}
\end{document}